\documentclass{PoS}
\usepackage{amsmath}
\usepackage{amssymb,xcolor,trimclip}
\usepackage{wrapfig}

\title{Early time dynamics and hard probes in heavy-ion collisions }

\ShortTitle{Early time dynamics and hard probes in heavy-ion collisions }

\author{\speaker{S.~Schlichting}\thanks{We thank A.~Kurkela, A.~Mazeliauskas, Y.~Mehtar-Tani, J.-F.~Paquet and D.~Teaney for insightful discussions and collaborations on the topics discussed in this proceeding. The author acknowledges support by the Deutsche Forschungsgemeinschaft (DFG, German Research Foundation) through the CRC-TR 211 'Strong-interaction matter under extreme conditions' -- project number 315477589 -- TRR 211.}\\
        Fakult\"{a}t f\"{u}r Physik, Universit\"{a}t Bielefeld, D-33615 Bielefeld, Germany\\
        E-mail: \email{sschlichting@physik.uni-bielefeld.de}}

\abstract{We present an overview of the current state of understanding of the early time dynamics of high-energy heavy-ion collisions, emphasizing recent developments and connections between the physics of the initial state and that of hard probes in heavy-ion collisions. Based on a weak-coupling description, we first establish a microscopic picture of the early time dynamics and equilibration process and subsequently discuss their macroscopic manifestations along with a novel macroscopic description of the pre-equilibrium phase. Some phenomenological consequences concerning the role of the pre-equilibrium phase in large (AA) and small (pp/pA) collision systems are briefly discussed along with open questions and opportunities for future improvements.}

\FullConference{International Conference on Hard and Electromagnetic Probes of High-Energy Nuclear Collisions\\
		30 September - 5 October 2018\\
		Aix-Les-Bains, Savoie, France}

\begin{document}

\section{Introduction}
High-energy heavy-ion collisions provide a unique environment to study the properties of strong-interaction matter under extreme conditions. Over the past decades, experiments at the Relativistic Heavy-Ion Collider (RHIC) and the Large Hadron Collider (LHC) have collected an overwhelming amount of evidence of the formation of a de-confined Quark-Gluon Plasma (QGP) , and established a standard picture of the space-time evolution of the QGP fireball produced in the collisions of heavy-nuclei. The study of the early time dynamics of high-energy collisions -- eventually leading to the formation of near-thermal QGP -- is a fascinating topic of research, which has led to a variety of interesting results concerning e.g. the range of applicability of viscous hydrodynamics or the emergence of universal properties in out-of-equilibrium systems. While theoretical studies have been performed for a long time to provide a theoretical justification for the applicability of dissipative fluid dynamics in high-energy heavy-ion collisions, the situation has changed considerably with recent experimental results hinting at the presence of a small droplet of QGP also in pA and pp collisions. Now the questions how exactly and under what conditions an almost equilibrated QGP is formed in hadronic collisions have come even more into focus, as they are expected to be of direct relevance for present and future experiments.

In this proceeding we provide a brief overview of the current state of understanding from a weak coupling perspective. We first establish the microscopic picture of the early time dynamics in Sec.~\ref{sec:micro}, emphasizing connections to the physics of hard probes in heavy-ion collisions. Subsequently, in Sec.~\ref{sec:macro} we highlight new theoretical developments \cite{Kurkela:2018vqr,Kurkela:2018wud} to establish a macroscopic description of the early time dynamics in heavy-ion collisions, based on linear response theory out-of-equilibrium. Some phenomenological consequences of the the early time dynamics are discussed in Sec.~\ref{sec:effects}, where we address the impact of the pre-equilibrium stage in nucleus-nucleus collisions and point out the importance of the early time non-equilibrium dynamics in the dynamical description of small systems realized e.g. in high-multiplicity proton-nucleus collisions ($p+A$). 
 
\section{Early time dynamics \& equilibration process -- a microscopic perspective}
\label{sec:micro}

\begin{figure}[t!]
\begin{center}
\includegraphics[width=0.8\textwidth]{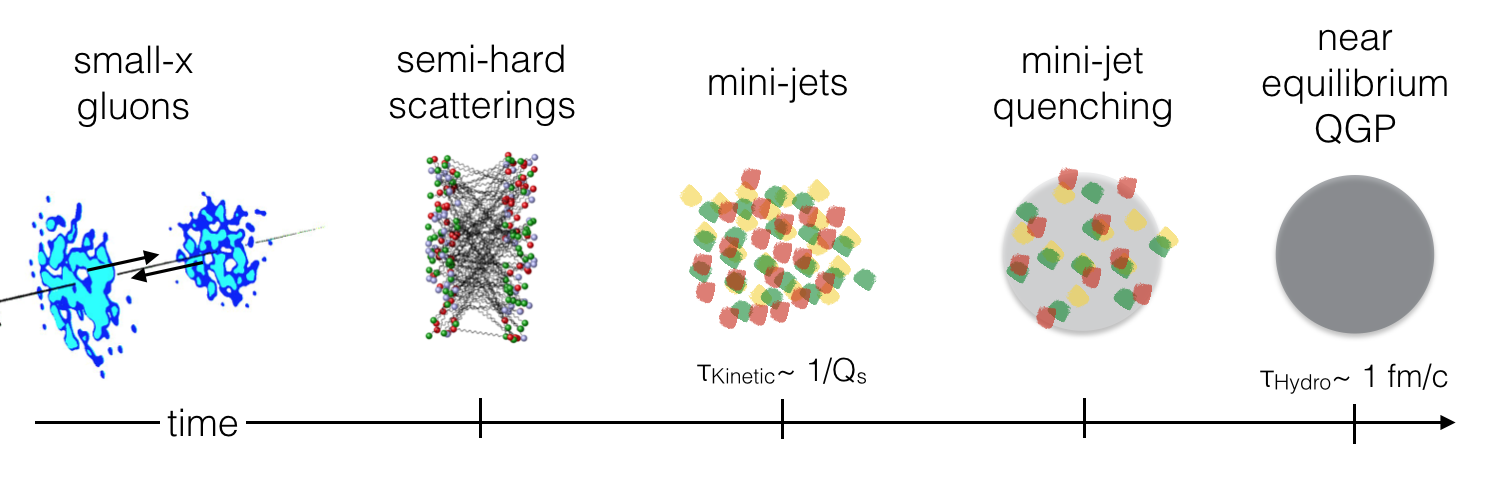}
\caption{\label{fig:cartoon} Illustration of different stages of the pre-equilibrium evolution in high-energy heavy-ion collisions. Beyond time scales $\tau_{\rm Kinetic} \sim 1/Q_s$ the approach towards local thermal equilibrium is described using QCD kinetic theory.}
\end{center}
\end{figure}

We will focus on the theoretical developments within a weak coupling description, which naturally connects the non-equilibrium dynamics at early times with that of hard probes, which remain out-of-equilibrium over the entire course of the space-time evolution. Based on a series of studies performed over the last few years \cite{Berges:2013eia,Berges:2013fga,Kurkela:2014tea,Kurkela:2015qoa} , a canonical picture of the early time dynamics in high-energy heavy-ion collision in the weak coupling limit has been established, which is compactly summarized in Fig.~\ref{fig:cartoon}. Starting with the partonic description of heavy nuclei, the non-equilibrium system of quarks and gluons produced in the collision undergoes a sequence of processes which eventually results in the formation of a close-to-equilibrium Quark-Gluon-Plasma (QGP). Based on this picture, the early time dynamics of particle production via multiple semi-hard scatterings is described in terms of classical field dynamics as incorporated in initial state models such as IP-Glasma\cite{Schenke:2012wb}. Beyond time scales $\tau \sim 1/Q_s$ the strong color fields lose their coherence, and the weakly coupled non-equilibrium plasma be efficiently described in terms of quasi-particles, with the characteristic transverse momentum of produced quarks and gluons is of the order of the saturation scale $Q_s$. Detailed studies of the classical field dynamics~\cite{Berges:2013eia,Berges:2013fga} indicate that the subsequent dynamics beyond $\tau \sim 1/Q_s$ can be efficiently described within QCD kinetic theory. 

\subsection{Effective kinetic description of equilibration process}
Based on the effective kinetic theory of QCD~\cite{Arnold:2002zm} the subsequent dynamics for $\tau \gtrsim 1/Q_s$ is described by a Boltzmann equation for the single particle phase-space density $f(x,p)$ of quarks and gluons\footnote{Note that present studies of the pre-equilibrium dynamics have focused on the Yang-Mills sector and do not include dynamical fermions in the kinetic description.}
\begin{eqnarray}
\label{eq:QCDkin}
p^{\mu} \partial_{\mu} f(x,p) = \mathcal{C}_{2\leftrightarrow 2}[f](x,p) + \mathcal{C}_{1\leftrightarrow 2} [f](x,p)
\end{eqnarray}
which at leading order weak coupling features two distinct processes associated with elastic $2\to2$ interactions, and inelastic nearly collinear (inverse) Bremsstrahlung processes which are effectively $1\leftrightarrow 2$  processes. Dynamical screening of $2\to2$ processes, and the Landau-Pomeranchuk-Migdal (LPM) suppression of the effective $1\leftrightarrow 2$ process are taken into account in terms of in-medium matrix elements~\cite{Arnold:2002zm}, which in practice are determined under the assumption of isotropic screening. Notably the basic theoretical framework for equilibration studies in Eq.~(\ref{eq:QCDkin}) is the same as in weakly coupled studies of parton/jet energy loss, implemented e.g. in the Monte-Carlo generators {\rm MARTINI} \cite{Schenke:2009gb} and {\rm JEWEL} \cite{Zapp:2013vla}. Nevertheless, there are also important differences in that for the typical degrees of freedom in the non-equilibrium plasma there is no clear hierarchy of scales as in jet physics where $E_{\rm jet} \gg T$, such that in practice all ``soft'' and ``(semi-) hard'' degrees of freedom have to be treated on equal footing, i.e. within the same kinetic framework, including also the mutual interactions between the typical (semi-) hard gluons.

\begin{figure}[t!]
\begin{center}
\includegraphics[width=0.8\textwidth]{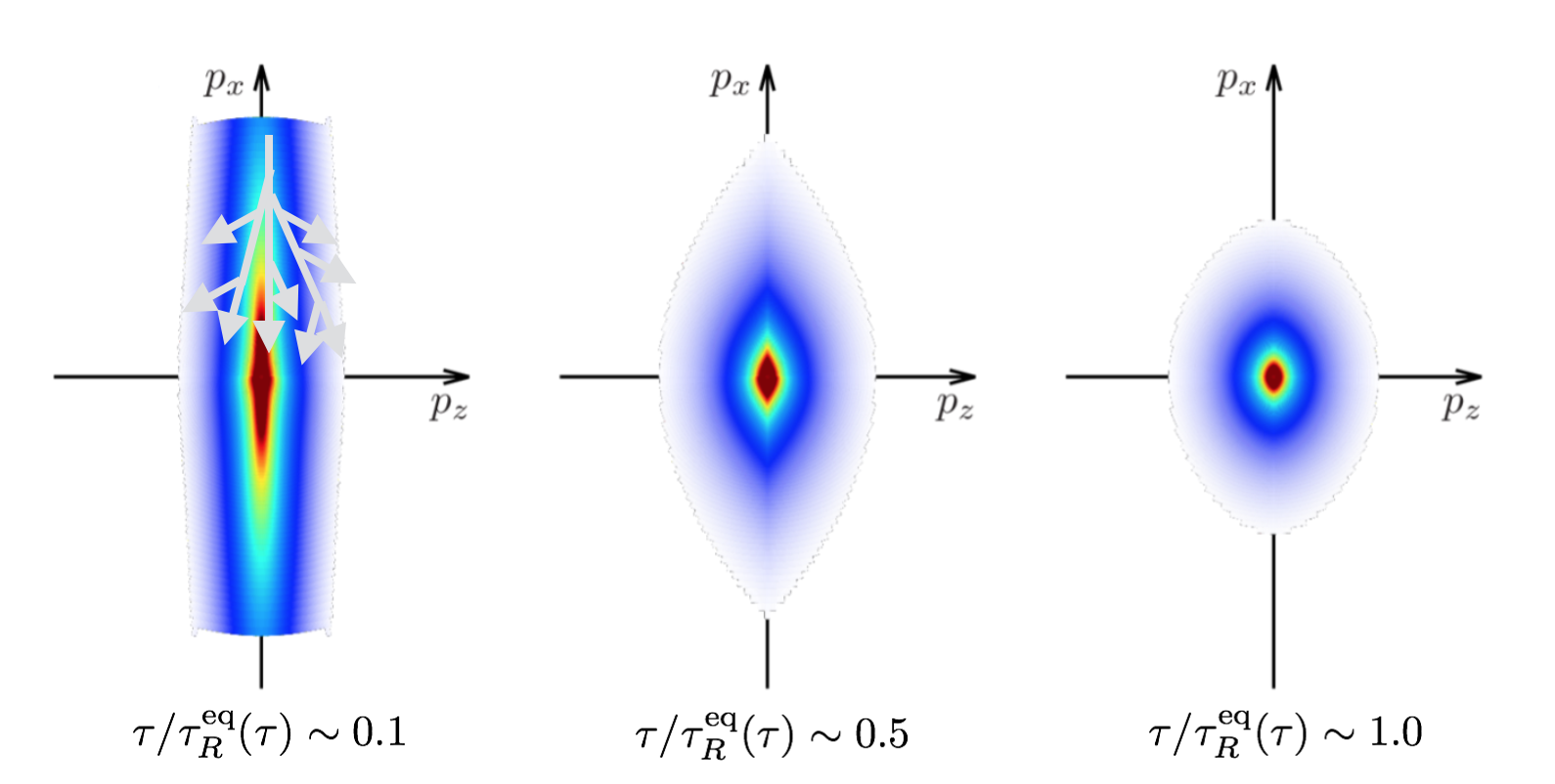}
\caption{\label{fig:bottomup} Kinetic theory simulation of the evolution of the phase-space distribution of gluons during the pre-equilibrium phase. \cite{Kurkela:2018vqr,Kurkela:2018wud} }
\end{center}
\end{figure}

\subsection{Mini-jet quenching \& bottom-up thermalization}
Basic microscopic picture of the subsequent equilibration process has been developed in a seminal paper by Baier et al.~\cite{Baier:2000sb}, along with first estimates of the relevant time scales. Numerical solutions of QCD kinetic theory \cite{Kurkela:2015qoa,Kurkela:2018vqr,Kurkela:2018wud}, confirm the qualitative picture of this ``bottom-up'' scenario, which is illustrated in Fig.~\ref{fig:bottomup}. Since the semi-hard gluons produced around mid-rapidity have much larger transverse momenta $p_{\bot}\sim Q_s$ than longitudinal momenta $p_{\|} \ll Q_s$ in the local rest-frame, the initial phase-space distribution shown in the left panel of Fig.~\ref{fig:bottomup} is highly anisotropic. Even though the strength of elastic scatterings is insufficient to isotropize the momentum distribution of semi-hard partons, the interactions between mini-jets cause additional Bremsstrahlung radiation, triggering a radiative break-up of the semi-hard $(p_{\bot} \sim Q_s)$ mini-jets into a large number of softer quanta $(p_{\bot} \ll Q_s)$. Since elastic interactions are significantly more efficient at lower momentum scales $(p_{\bot} \ll Q_s)$, these processes begin to isotropize the momentum distribution of the radiated quanta, eventually leading to the formation of a soft thermal bath seen in the central panel of Fig.~\ref{fig:bottomup}. In the final stages of the equilibration process, the remaining fragments of the semi-hard mini jets lose their energy to the soft thermal medium, very much like a jet/parton loosing energy to a thermal plasma. (Near-) equilibration of the system is completed once these mini-jets are fully quenched, i.e. the typical semi-hard partons with transverse momenta $p_{\bot} \sim Q_s$ have lost all their energy to the soft thermal bath. Since at weak coupling, the energy loss of higher momentum partons is approximately independent of their energy, the scale $Q_s$ also provides an estimate of the parton energy loss during the pre-equilibrium stage. However, a detailed investigation of parton/jet energy loss during the early stages of heavy-ion collisions is yet to be performed.

\subsection{Energy momentum tensor \& approach to hydrodynamics}
\begin{figure}[t!]
\begin{center}
\begin{minipage}{\textwidth}
\includegraphics[width=0.5\textwidth]{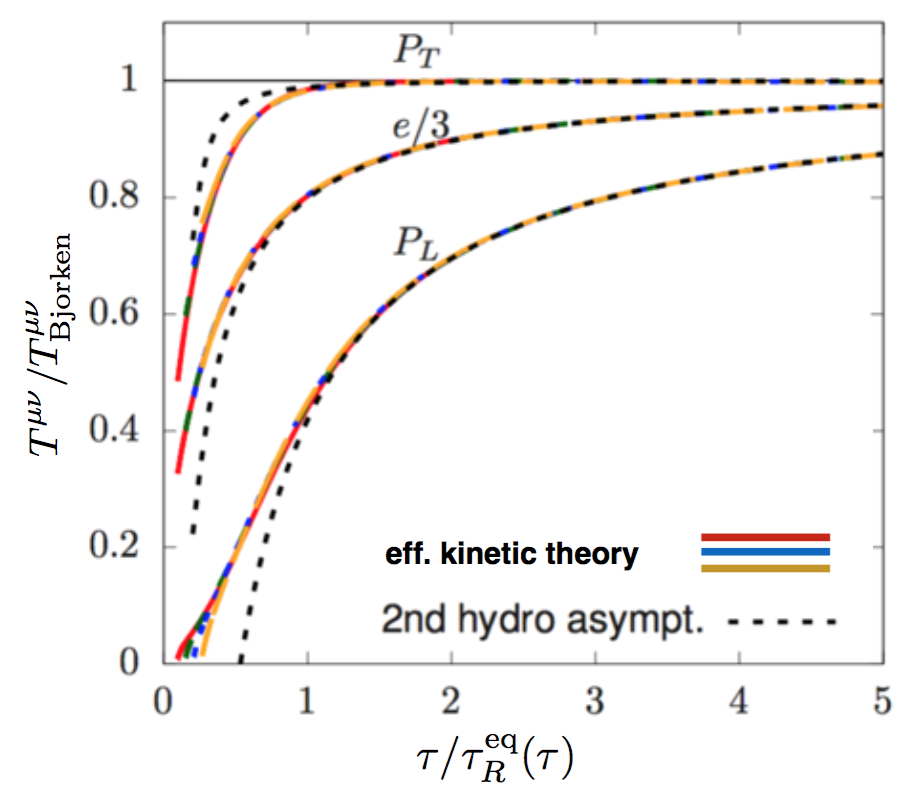}
\includegraphics[width=0.38\textwidth]{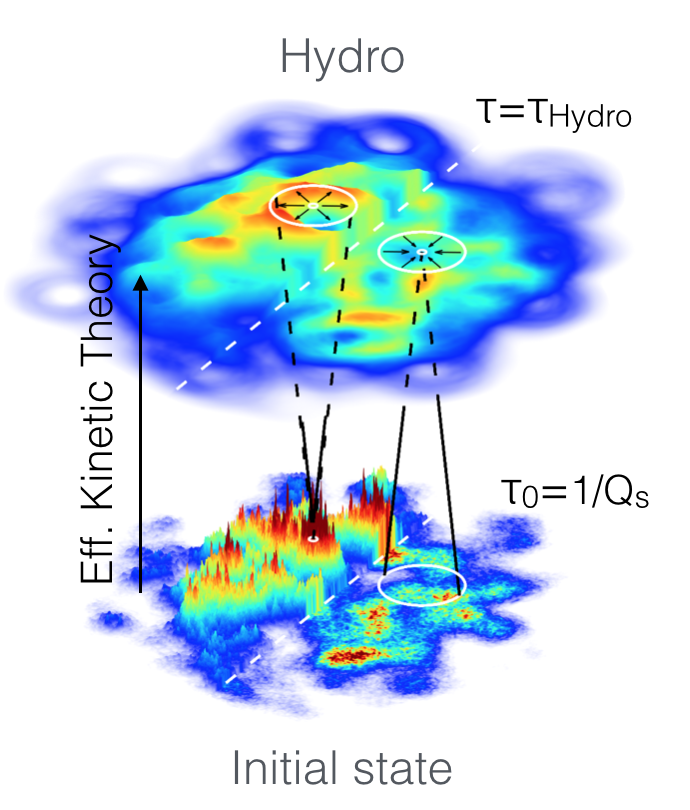}
\end{minipage}
\caption{\label{fig:energy} (left) Evolution of the energy-momentum tensor $T^{\mu\nu}=diag(e,P_T,P_T,P_L)$ during the pre-equilibrium stage. Different components $e/3,P_T,P_L$ are normalized to the asymptotic Bjorken value $T^{\mu\nu}_{\rm Bjorken}= \tau^{-4/3} \lim_{\tau \to \infty} \left(T^{\mu\nu} \tau^{4/3} \right)$ and plotted as a function of proper time $\tau/\tau_{R}^{\rm eq}(\tau)$, i.e. in units of the equilibrium relaxation time $\tau_{R}^{\rm eq}(\tau)=\frac{4\pi \eta/s}{T_{\rm Id}(\tau)}$ (see \cite{Kurkela:2018vqr} for details). Different colored curves corresponding to kinetic theory simulations at different coupling strength $\lambda=g^2N_c=5-20$ collapse onto a universal scaling curve. Beyond time scales $\tau/\tau_{R}^{\rm eq}(\tau)\gtrsim 1$ the non-equilibrium evolution follows the asymptotic behavior of second order hydrodynamics shown by dashed black curves shows. \cite{Kurkela:2018vqr,Kurkela:2018wud} (right) Illustration of macroscopic evolution of energy-momentum tensor during pre-equilibrium phase.  \cite{Kurkela:2018vqr,Kurkela:2018wud}  }
\end{center}
\end{figure}

Based on the microscopic picture described above, we will now proceed to describe the macroscopic evolution of the system during the pre-equilibrium stage. Since the system is initially highly anisotropic, with the longitudinal pressure $P_L$ much smaller than the transverse pressure $P_T$, one of the key questions is to understand how and on what time scales the energy momentum tensor $T^{\mu\nu}$ relaxes towards a local equilibrium state, where longitudinal and transverse pressure become identical $(P_L=P_T)$. Simulation results for the evolution of the different components of the energy momentum tensor $T^{\mu\nu}=diag(e,P_T,P_T,P_L)$ are presented in Fig.~\ref{fig:energy}, where different colored curves show the results for different coupling strength $\lambda=g^2N_c=5-20$. Strikingly, all results for different coupling strength collapse onto a universal scaling curve, once the evolution time $\tau$ is appropriately normalized in units of the respective equilibrium relaxation time $\tau_{R}^{\rm eq}(\tau)$, suggesting that -- in the relevant range of coupling strength -- the non-equilibrium evolution of the energy momentum tensor is in fact controlled by a single equilibrium time scale  
$\tau_{R}^{\rm eq}(\tau)= \frac{4 \pi \eta/s}{T_{\rm Id}(\tau)}$,
where $\eta/s$ is the shear-viscosity to entropy ratio and $T_{\rm Id}(\tau) \propto \tau^{-1/3}$ denotes an effective temperature of the non-equilibrium system.

Beyond time scales $\tau \gtrsim \tau_{R}^{\rm eq}(\tau)$, the different components of the energy momentum tensor start to follow viscous hydrodynamic behavior, allowing one to estimate 
the ``hydrodynamization'' time as  $\tau_{\rm Hydro}= 1 \rm{fm}/c \left(\frac{4\pi \eta/s}{2} \right)^{3/2} \left(  \frac{\langle s\tau \rangle}{ 4.1 {\rm GeV}^{2}} \right)$
where $\langle s\tau \rangle$ denotes the entropy density per unit rapidity -- a quantity that is tightly constrained from experimental measurements of the charged particle multiplicity ~\cite{Kurkela:2018vqr,Kurkela:2018wud}. Interestingly, one also observes from Fig.~\ref{fig:energy} that a viscous hydrodynamic description of the boost-invariant homogeneous system becomes applicable when the pressure anisotropy is still of $\mathcal{O}(1)$, with the longitudinal pressure $P_L$ significantly smaller than the transverse pressure $P_T$. While this may appear somewhat surprising at first glance, it is important to point out that a similar behavior has also been reported for strongly coupled systems \cite{Heller:2011ju,Casalderrey-Solana:2013aba} and is under active investigation in the context of non-equilibrium attractor solutions of dynamical systems \cite{Romatschke:2017vte,Blaizot:2017ucy,Strickland:2017kux}.

\section{Macroscopic description of early time dynamics}
\label{sec:macro}
So far we have discussed the dynamics of the pre-equilibrium stage based on an effective kinetic description of the phase-space densities $f(x,p)$ of microscopic degrees of freedom of the plasma. However, from the point of view of a phenomenological description of high-energy heavy-ion collisions, one is primarily interested in the macroscopic features, such as the event-by-event profiles of the energy momentum tensor $T^{\mu\nu}(x)$ at some initial time $\tau_{\rm Hydro}\sim 1 {\rm fm}/c$ where a fluid dynamical description is applicable. Since many of of the microscopic features become irrelevant as the system approaches a state of local thermal equilibrium, it is suggestive that the evolution of the energy momentum tensor can readily be obtained from a coarse-grained description of the pre-equilibrium stage \cite{Keegan:2016cpi}.

\begin{figure}[t!]
\begin{center}
\includegraphics[width=\textwidth]{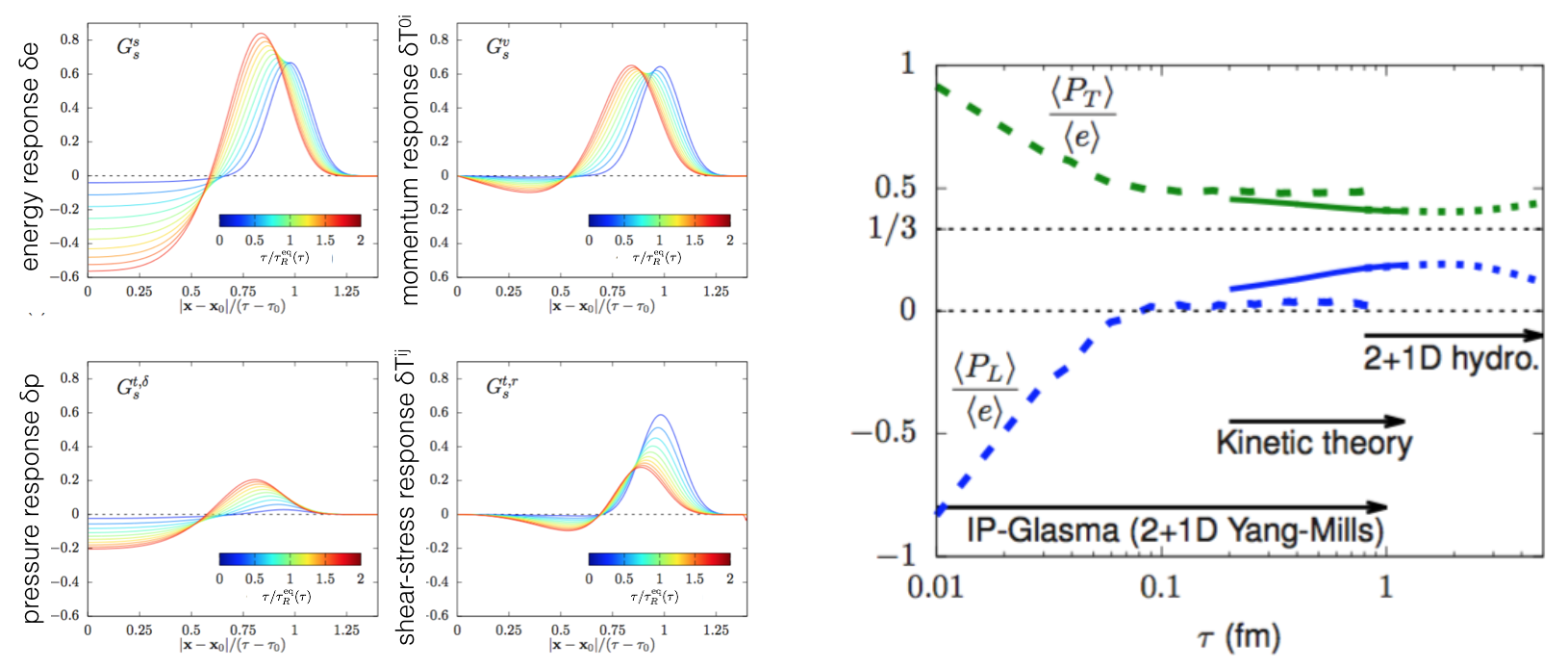}
\caption{\label{fig:four} (left) Non-equilibrium Green's functions $G^{\mu\nu}_{\tau\tau}$ in Eq.~(\ref{eq:linrespo}) characterizing the response $\delta T^{\mu\nu}(\tau)$ of the various components  the energy-momentum tensor to an initial energy perturbation $\delta T^{\tau\tau}(\tau_0)$.  \cite{Kurkela:2018vqr,Kurkela:2018wud} (right) Evolution of long. and transverse pressures $P_T$ and $P_L$ in a central $Pb+Pb$ collision -- a consistent description of the dynamics from early to late times is obtained by matching different theoretical descriptions.  \cite{Kurkela:2018vqr,Kurkela:2018wud}}
\end{center}
\end{figure}

New developments in this direction have been put forward in \cite{Kurkela:2018vqr,Kurkela:2018wud}, which established a macroscopic description of the pre-equilibrium evolution of the energy-momentum tensor $T^{\mu\nu}$, based on a separation of scales between the causal propagation radius $c (\tau_{\rm Hydro}- \tau_0)$, which is much smaller than the characteristic size scale of the relevant gradients $\sim 1/R_A$. Since on the scales $c (\tau_{\rm Hydro}- \tau_0)$ of the causal horizon illustrated in the right panel of Fig.~\ref{fig:energy}, fluctuations $\delta T^{\mu\nu}(x)$ of the energy-momentum tensor are typically small compared to the local average $T^{\mu\nu}_{BG}(x)$, the non-equilibrium system can be treated as approximately homogeneous, with small deviations described in linear response theory. Exploiting also the fact that, there is an effective memory loss during the pre-equilibrium stage, such that the final energy momentum tensor $T^{\mu\nu}(\tau_{\rm Hydro})$ is insensitive to the microscopic details of the initial phase-space distribution $f(\tau_0)$, the energy-momentum tensor at $\tau=\tau_{\rm Hydro}$ can then be directly constructed from a given set of initial conditions at initial time $\tau_0$ according  to
\begin{eqnarray}
\label{eq:linrespo}
T^{\mu\nu}(\tau,{\bf x}) \simeq \underbrace{T^{\mu\nu}_{\rm BG,{\bf x}}(\tau)}_{\substack{\text{non-eq. evolution of} \\ \text{(local) avg. background}}} + \underbrace{\int_{\odot} d^2{\bf x}_{0}~G^{\mu\nu}_{\alpha\beta}(\tau,\tau_0,{\bf x},{\bf x}_0)~\delta T^{\alpha\beta}(\tau_0,{\bf x}_0) }_{\text{non-eq. evolution of local fluctuations of energy-momentum tensor}} + \cdots\;,
\end{eqnarray}
where $\cdots$ indicate additional contributions due non-linear effects and higher order moments (of $f(x,p)$), which have been neglected in \cite{Kurkela:2018vqr,Kurkela:2018wud}. Clearly, the advantage of this formulation is that the microscopic properties of the system only enter into the calculation of the average energy momentum tensor $T^{\mu\nu}_{\rm BG,{\bf x}}(\tau)$ (shown in Fig.~\ref{fig:four}) and the non-equilibrium response functions $G^{\mu\nu}_{\alpha\beta}(\tau,\tau_0,{\bf x},{\bf x}_0)$ (shown in the left panel of Fig.~\ref{fig:four}). Based on the knowledge of these two quantities, which have been calculated in QCD kinetic theory~\cite{Kurkela:2018vqr,Kurkela:2018wud} and are publicly available as part of the KoMP{\o}ST package,\footnote{KoMP{\o}ST: https://github.com/KMPST/KoMPoST} the pre-equilibrium evolution of the energy momentum can be directly obtained from Eq.~\ref{eq:linrespo}, providing an interesting alternative to performing event-by-event Monte-Carlo calculations of the microscopic dynamics. It is worth point out, that such a macroscopic formulation in terms linear response functions could also be interesting to explore in the context of integrated jet+medium studies to macroscopically describe the energy deposition due to thermalization of soft jet-fragments. 

Most importantly, the macroscopic description in Eq.~(\ref{eq:linrespo}) can be implemented in a straightforward way to include an event-by-event description of the pre-equilibrium stage into hydrodynamic simulations of the space-time evolution of heavy-ion collisions. A proof of principle calculation in this respect is depicted in the right panel of Fig.~\ref{fig:four}, where the energy momentum tensor from classical Yang-Mills simulations of particle production at early times $\tau\lesssim 0.2 {\rm fm/c}$ has been used as an input to the macroscopic kinetic description of the equilibration process, which in turn provides the initial conditions for subsequent viscous hydrodynamic evolution of the fireball beyond time scales $\tau\gtrsim 1.0 {\rm fm/c}$.  Strikingly, the overlap in the range of validity of the different descriptions \cite{Berges:2013fga,Berges:2013eia} ensures that there is essentially no sensitivity to variations of the matching times, as discussed in detail in \cite{Kurkela:2018vqr,Kurkela:2018wud}.

\section{Effects of the pre-equilibrium dynamics in small \& large system}
\label{sec:effects}
Based on an improved theoretical understanding of the early time non-equilibrium dynamics in high-energy collisions, it is interesting to explore some of the phenomenological consequences. So far no systematic studies have been performed in this respect, however one striking observation concerns the rapid production of entropy, which increases by approximately a factor of $2-3$ during the kinetic phase. Since there is essentially a one to one correspondence between the entropy density and charged particle multiplicity, it is clear that this should be taken into account when estimating e.g. charged particle multiplicities from initial state models as e.g. in \cite{Abreu:2007kv}, or vice versa when trying to constrain the properties of the initial state from experimental data.

Notably, the improved understanding of the early time dynamics and approach to equilibrium can also be used to constrain the space-time dynamics of small systems created in high multiplicity proton-proton and proton-nucleus collisions. By comparing the characteristic time scale $\tau_{\rm Hydro}$ for the formation of a near-equilibrium QGP to the system size $R$, one concludes that for $\tau_{\rm Hydro} \gg R$ one expects almost no final state interactions, as the system begins to fall apart before local equilibrium can be reached. Conversely, for $\tau_{\rm Hydro} \ll R$ the quark-gluon matter produced in the collision equilibrates quickly and subsequently experiences a long lived hydrodynamic phase. Estimates of the relevant ratio 
$\frac{\tau_{\rm Hydro}}{R} \simeq \left( \frac{\eta/s}{ 2/(4\pi)}\right)^{3/2} \left(  \frac{dN{\rm ch}/d\eta}{63}\right)^{-1/2} \left( \frac{S/N_{\rm ch}}{7}\right)^{-1/2}$ \cite{Kurkela:2018vqr,Kurkela:2018wud} show that the experimental results in pp/pA collisions (where $dN_{\rm ch}/d\eta \sim 10 -100$) mostly fall into the regime where $\tau_{\rm Hydro}/R \sim 1$, whereas in central and mid-central AA collisions  (where $dN_{\rm ch}/d\eta \sim 1000$) one has $\tau_{\rm Hydro} / R \ll 1$, indicating that a genuine non-equilibrium description will be needed to describe the dynamics of small systems across the wide range of multiplicities explored in experiments. While first developments in this direction have been reported in the literature based on partonic transport models \cite{Greif:2017bnr} \& semi-analytic transport studies \cite{Romatschke:2018wgi,Kurkela:2018ygx,Borghini:2018xum}, a comprehensive description of the out-of-equilibrium dynamics from QCD remains an outstanding challenge.

Based on the microscopic picture of the equilibration process discussed in Sec.~\ref{sec:micro}, one can further expect -- at least on a qualitative level  -- that signatures for the onset of thermalization of the bulk matter should be accompanied by quenching of hard probes in small systems. Despite all the complications associated with the study of hard probes in small systems, experiments are starting to provide stringent constrains, e.g. on the energy loss of reconstructed jets \cite{Acharya:2017okq}, and preliminary studies based on fluid-dynamical + jet simulations \cite{Shen:2016egw,Park:2016jap}, hint at sizable effects in high-multiplicity events. However, there are currently no dedicated theoretical studies combining the out-of-equilibrium dynamics of soft and hard degrees of freedom and the key question, to what extent hard probes can serve as control measurement to constrain the bulk dynamics in small systems will require further investigations in the future.

\section{Conclusions \& Outlook}
Significant progress in understanding the early time dynamics of high-energy collisions has enabled a complete description of the pre-equilibrium dynamics of heavy-ion collisions based on a combination of weak coupling techniques. While on the microscopic level the various stages of the equilibration process share important features with the dynamics of parton/jet energy loss, one of the most important developments comes from the macroscopic description of the pre-equilibrium dynamics provides, which enables for the first time an event-by-event description of the out-of-equilibrium space-time dynamics. So far studies of the early time dynamics have primarily focused on the kinetic equilibration of the QGP; however within the same framework it will also be possible to study e.g. the dynamics of quark production/chemical equilibration or the behavior of electro-magnetic and hard probes of the QGP.

\end{document}